\documentclass[conference]{IEEEtran}
\IEEEoverridecommandlockouts
\usepackage{amsmath}
\usepackage[caption=false]{subfig}
\usepackage{tabularx,graphicx}
\usepackage{hyperref}
\usepackage[numbers,sort&compress]{natbib}

\begin{document}

\title{The Evaluation of Rating Systems\\in Online Free-for-All Games}

\author{\IEEEauthorblockN{Arman Dehpanah}
\IEEEauthorblockA{\textit{School of Computing} \\
\textit{DePaul University}\\
Chicago, USA \\
\small{adehpana@depaul.edu}}
\and
\IEEEauthorblockN{Muheeb Faizan Ghori}
\IEEEauthorblockA{\textit{School of Computing} \\
\textit{DePaul University}\\
Chicago, USA \\
\small{mghori2@depaul.edu}}
\and
\IEEEauthorblockN{Jonathan Gemmell}
\IEEEauthorblockA{\textit{School of Computing} \\
\textit{DePaul University}\\
Chicago, USA \\
\small{jgemmell@cdm.depaul.edu}}
\and
\IEEEauthorblockN{Bamshad Mobasher}
\IEEEauthorblockA{\textit{School of Computing} \\
\textit{DePaul University}\\
Chicago, USA \\
\small{mobasher@cs.depaul.edu}}
}
 
\maketitle

\begin{abstract}

Online competitive games have become increasingly popular.
To ensure an exciting and competitive environment, these games routinely attempt to match players with similar skill levels.
Matching players is often accomplished through a rating system.
There has been an increasing amount of research on developing such rating systems.
However, less attention has been given to the evaluation metrics of these systems.
In this paper, we present an exhaustive analysis of six metrics for evaluating rating systems in online competitive games.
We compare traditional metrics such as accuracy.
We then introduce other metrics adapted from the field of information retrieval.
We evaluate these metrics against several well-known rating systems on a large real-world dataset of over 100,000 free-for-all matches.
Our results show stark differences in their utility.
Some metrics do not consider deviations between two ranks.
Others are inordinately impacted by new players.
Many do not capture the importance of distinguishing between errors in higher ranks and lower ranks.
Among all metrics studied, we recommend Normalized Discounted Cumulative Gain (NDCG) because not only does it resolve the issues faced by other metrics, but it also offers flexibility to adjust the evaluations based on the goals of the system.

\end{abstract}


\section{Introduction}

Online competitive games pit players against one another in player-versus-player (PvP) matches.
A common goal of PvP games is to match players based on their skills.
When a new player is matched against an experienced player, neither is likely to enjoy the competition.
Therefore, competitive games often use rating algorithms to match players with similar skills.

Rating systems often represent players with a single number, describing the player's skills.
For example, Elo~\cite{elo1978rating} considers 1500 as the default skill rating for new players and updates this value based on the outcome of the matches they played.
Rating systems leverage skill ratings to predict ranks.
While researchers have made numerous efforts in improving rank prediction, less attention has been given to how predicted ranks are evaluated.

There are several metrics commonly used to evaluate ratings.
However, these metrics often do not capture important characteristics of the ratings.
They might give equal weighting to low-tier and top-tier players, even when matching top-tier players is more important for the goals of the system.
They might also be hampered by the inclusion of new players since the system does not possess any knowledge of these players.

In this paper, we consider six evaluation metrics.
We include traditional metrics such as accuracy, mean absolute error, and Kendall's rank correlation coefficient.
We further include metrics adapted from the domain of information retrieval, including mean reciprocal rank (MRR), average precision (AP), and normalized discounted cumulative gain (NDCG).
We analyze the ability of these metrics to capture meaningful insights when they are used to evaluate the performance of three popular rating systems: Elo, Glicko, and TrueSkill.

To perform this analysis we limit our experimentation to free-for-all matches.
Free-for-all is a widely used game-play mode where several players simultaneously compete against one another in the same match.
The winner is the ``last man standing''; this mode of game-play is more commonly referred to as Battle Royale.
Our real-world dataset includes over 100,000 matches and over 2,000,000 unique players from PlayerUnknown's Battlegrounds.

Our evaluation shows that in free-for-all matches the metrics adapted from information retrieval can better evaluate the rating systems while being more resistant to the influence of new players.
NDCG, in particular, could more precisely capture the predictive power of these systems.
NDCG distinguishes between the prediction errors for top-tier players with higher ranks and those for low-tier players with lower ranks by applying a weight to positions.
This is particularly important for companies whose business model is based on user interaction and engagement at the top levels of play.
Top-tier players are the ones who most probably stay in the system and play more games.

The rest of this paper is organized as follows:
In Section~\ref{sec:related}, the related works are reviewed.
Rank prediction and its application in rating systems are discussed in Section~\ref{sec: prediction}.
In Section~\ref{sec: metrics}, the evaluation metrics are introduced.
In Section~\ref{sec: methodology}, the dataset and the experiments are explained, and then we discuss the results in detail in Section~\ref{sec: results}.
Finally, we conclude the paper and mention future works in Section~\ref{sec: conclusion}.


\section{Related Work}
\label{sec:related}

The popularity of online competitive games has exploded in the last decade.
There are more than 800 million users playing online games and this number is expected to grow to over 1 billion by the year 2024~\cite{website1}.

An important mode of game-play is the free-for-all which can be divided into deathmatch and battle royale.
In deathmatch games many players are pitted against one another; the winner is the one with the most points at the end of the game.
In battle royale players eliminate one another; the winner is the last one standing.

A critical component of these zero-sum games is the ability to rate a player's skill.
An accurate skill rating enables the system to generate balanced matches~\cite{myslak2014developing, herbrich2007trueskill}.
It allows for the assignment of teams with similar skill levels~\cite{zhang2010factor,delalleau2012beyond,myslak2014developing,menke2008bradley,buckley2013predicting}.
It serves as feedback for the players so that they can track their performance over time~\cite{herbrich2007trueskill,buckley2013predicting,lopez2014continuous}.

The most prominent examples of such systems are Elo~\cite{elo1978rating}, Glicko~\cite{glickman1999parameter} and TrueSkill~\cite{herbrich2007trueskill}.
Elo and Glicko are designed for head-to-head matches with only two players.
TrueSkill, developed by Microsoft, extends upon these approaches to handle multiplayer games with multiple teams.

A rating system like those above can be used to predict the outcome of games.
Those predictions can then be evaluated to judge the quality of the rating system.
Accuracy is often used for evaluating rating systems.
It has been used to evaluate first-person shooter games~\cite{herbrich2007trueskill, menke2008bradley, zhang2010factor, delalleau2012beyond, delong2011teamskill, makarov2017predicting, weng2011bayesian, guo2012score}, real-time strategy games~\cite{zhang2010factor, makarov2017predicting, chen2016predicting, li2018learning}, as well as tennis\cite{chen2016predicting, motegi2012network, ibstedt2019application}, soccer~\cite{guo2012score, ibstedt2019application}, football~\cite{guo2012score}, and board games~\cite{cooper2016player, morrison2019comparing, stanescu2011rating}.

Other metrics have been used for rank prediction.
Log-likelihood has been used to evaluate first-person shooters~\cite{delalleau2012beyond}, real-time strategies~\cite{chen2016predicting}, and board games~\cite{stanescu2011rating}.
Information gain has been used to evaluate first-person shooters~\cite{guo2012score}.
Mean squared error has been used to evaluate real-time strategy games~\cite{yu2018moba} and soccer~\cite{lasek2013predictive}.
Mean absolute error has been used to evaluate first-person shooters~\cite{guo2012score} and real-time strategies~\cite{yu2018moba}.
Root mean squared error has been used to evaluate chess games~\cite{morrison2019comparing}.
Spearman's rank correlation coefficient has been used to evaluate first-person shooter games~\cite{buckley2015rapid}.
However, most of these examples are head-to-head games; there are only two sides such as in chess or squad-vs-squad first-person shooters.
When there are more than two sides, these metrics are less appropriate.

Information retrieval, the process of obtaining relevant resources from document collections~\cite{manning2008introduction}, includes several metrics for evaluating the ranking of search results.
These metrics can be adapted to the evaluation of predictions in online competitive games.
Mean reciprocal rank considers the rank positions of relevant documents in the results list to compute relevance scores~\cite{voorhees1999trec}.
Average precision considers the number of relevant documents among retrieved documents along with the number of documents retrieved out of all available relevant documents to evaluate the performance of the system~\cite{salton1991developments}.
NDCG considers a graded relevance for determining the gain obtained by each retrieved document~\cite{jarvelin2002cumulated}.

Our work differs from previous efforts.
We focus on free-for-all games, one of the most popular game-play modes in which several players compete against one another in a single match.
We extend Elo and Glicko to free-for-all games.
Whereas research often attempts to improve rank prediction, we seek to better evaluate the predicted ranks.
We explore six evaluation metrics, including traditional metrics and those drawn from the domain of information retrieval.
We analyze the explanatory power of these metrics using a large real-world dataset and consider how the metrics describe different populations such as the top-tier and the most frequent players.


\section{Predicting Rank}
\label{sec: prediction}

Predicting rank in online competitive games is important for several reasons.
It can be used to evaluate the performance of the players.
It can be used to assign players to teams.
It can be used to create balanced matches.
Often, rank is predicted by first evaluating the skills of the players.

The skill level of a player $p$ can be described as a single number: $\mu_p$.
In a head-to-head match between two players, $p_1$ and $p_2$, with skill ratings, $\mu_{p_1}$ and $\mu_{p_2}$, the player with the higher rating can be predicted to win that match.
In a free-for-all match with several players $p_1, p_2, p_3, ..., p_n$, their skill ratings $\mu_{p_1}, \mu_{p_2}, ..., \mu_{p_n}$ can be used to create a rank ordering of the players $R^{pred}$.
After the match, the observed rankings $R^{obs}$ can be used to update the skill levels.

There are several ways to update a player's skill rating.
In the remainder of this section, we describe three common algorithms: Elo, Glicko, and TrueSkill.


\subsection{Elo}
\label{sec:elo}

Originally developed for ranking chess players, Elo has been used for ranking players in many competitive environments~\cite{elo1978rating}.
Elo calculates the skill level of players by appraising a set of historical match results. 
Elo assumes that players' skill follows a Gaussian distribution with the same standard deviation for all players.

As a convention, the default rating for new players is set to 1500.
The rating is updated after each match.
The winner is awarded points and the loser surrenders points after each match.
The amount of these points is dependent on the probability of the outcome of the match based on the two players' initial ratings.

The probability that player $p_i$ wins the match against player $p_j$ can be calculated by:

\setlength{\belowdisplayskip}{6pt} \setlength{\belowdisplayshortskip}{6pt}
\setlength{\abovedisplayskip}{2pt} \setlength{\abovedisplayshortskip}{2pt}

\small
\begin{equation*}
    Pr(p_i\:wins, p_j) = \big(1 + e^{\frac{\mu_j - \mu_i}{D}}\big)^{-1}
\end{equation*}
\normalsize

\noindent where $\mu_i$ and $\mu_j$ are the ratings of the two players.
\textit{D} represents the weight given to ratings when determining players' estimated scores.
Using higher values for \textit{D} decreases the influence of the difference between ratings and vice versa.
Conventionally, \textit{D} is set to 400.

After the match, the rating of player $p_i$ is updated by:

\small
\begin{equation*}
\label{eqn:elo_update}
    \mu_i^\prime = \mu_i + K[R - Pr(p_i\: wins, p_j)]
\end{equation*}
\normalsize

\noindent where $R$ is 1 if player $p_i$ wins the game, 0.5 if the match is a draw, and 0 if it is a loss.
\textit{K} is a scaling factor determining the magnitude of the change to players' ratings after each match.
Using higher values for \textit{K} leads to greater changes in players' ratings.
The value of \textit{K} should be tuned based on the nature of the game and the players' characteristics.
For example, the World Chess Federation (FIDE) considers several tiers for the value of \textit{K}.
It uses \textit{K = 40} for new players until they participate in at least 30 matches, \textit{K = 20} as long as their ratings remain under 2400, and \textit{K = 10} once their ratings reach 2400.

We can extend Elo skill ratings from head-to-head matches to free-for-all matches that include many players.
Several possibilities exist.
One way is to consider the match as a set of head-to-head matches.

We recalculate the probability of winning for each player by summing the probability of winning values in all their pairwise matches versus the other players.
Assuming \textit{N} players competing against each other in a field \textit{F}, the overall probability of winning for player $p_i$ can be calculated as:

\small
\begin{equation*}
    Pr(p_i\:wins, F) = \frac{\sum\limits_{1 \leq j \leq N, i \neq j }  \big({1 + e^{\frac{(\mu_j - \mu_i)}{D}}}\big)^{-1}}{{\binom{N}{2}}} 
\end{equation*}
\normalsize

\noindent where $\binom{N}{2}$, the total number of pairwise comparisons, is used to normalize the probability values to sum up to 1.

Because free-for-all matches include several players, we normalize the sum of the observed outcomes to 1 in order to conform to Elo's design that the total number of points awarded is equal to the total number of points deducted.
For player $p_i$ we transform the observed rank $R^{obs}_i$ into a normalized result, $R^{'}_{i}$, calculated as:

\small
\begin{equation*}
\label{eqn:R}
    R^{'}_{i} = \frac{N - R^{obs}_i}{\binom{N}{2}}
\end{equation*}
\normalsize

The player's Elo rating in this multi-player environment can then be updated as:

\small
\begin{equation*}
    \mu_i^\prime = \mu_i + K[R^{'}_{i} - Pr(p_i\:wins, F)]
\end{equation*}
\normalsize

One criticism of Elo is that it assumes a fixed skill variance for all players and may not handle uncertainty well.
This could result in reliability issues~\cite{glickman1999parameter}.
Glicko addresses this problem.


\subsection{Glicko}
\label{sec:glicko}

The Glicko rating system~\cite{glickman1999parameter} extended Elo by introducing a dynamic skill deviation $\sigma$ for each player.
Players are characterized by a distribution with a mean $\mu$ representing their skill and a deviation $\sigma$ representing the uncertainty about their skill.
The frequency that a player competes in the game is used to modify their skill deviation $\sigma$.
New players are assigned $\mu = 1500$ and $\sigma = 350$.
Both these numbers are updated after each match.

The probability that player $p_i$ wins the match against player $p_j$ can be calculated by:

\small
\begin{equation*}
    Pr(p_i\:wins, p_j) = \big(1 + 10^{\frac{-g(\sqrt{\sigma_i^2 + \sigma_j^2})(\mu_i - \mu_j)}{400}}\big)^{-1}
\end{equation*}
\normalsize

\noindent where $\mu_i$, $\mu_j$, $\sigma_i$, and $\sigma_j$ represent the skill ratings and skill deviations of the two players.
The function \textit{g} takes the sum of the square of the two skill deviations and uses them to weight the deviation in the players' skills. It is defined as:

\small
\begin{equation*}
    g(\sigma) = \big(\sqrt{\frac{1 + 3q^2\sigma^2}{\pi^2}}\big)^{-1}
\end{equation*}
\normalsize

\noindent where Glicko sets \textit{q} as a constant equal to $0.0057565$.
After the match, the skill rating and deviation of player $p_i$ are updated:

\small
\begin{equation*}
    \mu_i^{\prime} = \mu_i + \frac{q}{\frac{1}{\sigma_i^2} + \frac{1}{d^2}} \big[ g(\sigma_j)(R - Pr(p_i\:wins, p_j)) \big]
\end{equation*}

\begin{equation*}
    \sigma_i^{\prime} = \sqrt{\big(\frac{1}{\sigma_i^2} + \frac{1}{d^2}\big)^{-1}}
\end{equation*}
\normalsize

\noindent where \textit{R} is 1 if player $p_i$ wins the game, 0.5 if the match is a draw, and 0 if it is a loss.
The variable $d^2$ is minus of inverse of Hessian of the log marginal likelihood and is calculated as:

\small
\begin{equation*}
    d^2 = \big[ q^2 g(\sigma_j)^2 Pr(p_i\:wins, p_j)(1- Pr(p_i\:wins, p_j)) \big]^{-1}
\end{equation*}
\normalsize

We can extend Glicko to free-for-all matches as we did with Elo.
We consider each free-for-all match with \textit{N} players as $\binom{N}{2}$ separate matches between each pair of players.
The probability of winning for player $p_i$ can be calculated as:

\small
\begin{equation*}
    Pr(p_i\:wins, F) = \frac{\sum\limits_{1 \leq j \leq N, i \neq j } \big({1 + 10^{\frac{-g(\sqrt{\sigma_i^2 + \sigma_j^2})(\mu_i - \mu_j)}{400}}}\big)^{-1}}{\binom{N}{2}}
\end{equation*}
\normalsize

Again we normalize the function with $\binom{N}{2}$ so that the probability values sum up to 1.
The variable $d^2$ can be updated for this scenario as:

\small
\begin{equation*}
    d^2 = \big[ q^2 g(\sigma_j)^2 Pr(p_i\:wins, F)(1- Pr(p_i\:wins, F)) \big]^{-1}
\end{equation*}
\normalsize

Similar to Elo, Glicko is a zero-sum rating system; an equal number of points is awarded and deducted.
In order to achieve such a balance in a multi-player free-for-all match, we once again normalize the match results as before.
Finally, the rating of player $p_i$ can be updated as:

\small
\begin{equation*}
    \mu_i^{\prime} = \mu_i + \frac{q}{\frac{1}{\sigma_i^2} + \frac{1}{d^2}} \big[ g(\sigma_j)(R^{'}_{i} -  Pr(p_i\:wins, F))\big] 
\end{equation*}
\normalsize

Glicko has been proven to be successful at rating players.
However, it does have some drawbacks.
Glicko requires an average of 5 to 10 matches for each player in order to accurately describe a player's skill~\cite{glickman1995glicko}.
Moreover, when players compete very frequently, their skill deviation $\sigma$ becomes very small and there are no noticeable changes in their ratings, even when they are truly improving.
Finally, while we have extended Elo and Glicko from head-to-head matches to large multiplayer free-for-all matches, they were not initially designed to do so.
TrueSkill was designed for this purpose.


\subsection{TrueSkill}
\label{sec:trueskill}

TrueSkill~\cite{herbrich2007trueskill} is a Bayesian ranking system developed by Microsoft Research for Xbox Live that can be applied to any type of game-play mode with any number of players or teams.
TrueSkill derives individual skill levels from the outcome of matches between players by leveraging factor graphs~\cite{kschischang2001factor} and expectation propagation algorithm~\cite{minka2001family}. 

Similar to Glicko, TrueSkill assumes that the performance of players follows a Gaussian distribution with mean $\mu$ and standard deviation $\sigma$ representing their skills and skill deviations.
New players are assigned $\mu = 25$ and $\sigma = 8.333$. 
These values are updated after each match.

TrueSkill follows different update methods depending on whether a draw is possible.
For a non-draw case, if $\mu_i$, $\mu_j$, $\sigma_i$, and $\sigma_j$ represent skill ratings and deviations of players $p_i$ and $p_j$, assuming player $p_i$ wins the match against player $p_j$, his skill rating is updated by:

\small
\begin{equation*}
    \mu_i^{\prime} = \mu_i + \frac{\sigma^{2}_i}{c}\big[ \frac{N(\frac{t}{c})}{\Phi(\frac{t}{c})}\big]
\end{equation*}
\normalsize

\noindent where $t = \mu_i - \mu_j$ and $c = \sqrt{2\beta^2 + \sigma_i^2 + \sigma_j^2}$. 
$N$ and $\Phi$ represent the probability density function and cumulative distribution function of a standard normal distribution.
The parameter $\beta$ is the scaling factor determining the magnitude of changes to ratings.
Skill deviations for both players are updated by:

\small
\begin{equation*}
    \sigma^{\prime} =  \sigma - \sigma \big(\frac{\sigma^2}{c^2}  \big[\frac{N(\frac{t}{c})}{\Phi(\frac{t}{c})}\big]    \big[\frac{N(\frac{t}{c})}{\Phi(\frac{t}{c})} + t\big] \big)
\end{equation*}
\normalsize

TrueSkill has been used in online games~\cite{huang2013mastering}, sports~\cite{ibstedt2019application}, education~\cite{kawatsu2017predicting}, recommender systems~\cite{quispe2015content}, and click prediction for online advertisements~\cite{graepel2010web}.  
Despite the popularity of TrueSkill, it suffers from a conceptual issue.
It ignores interactions of players within a team and assumes their performance is independent of one another.
This issue was addressed by several following works through which many different algorithms and extensions were introduced~\cite{delalleau2012beyond, guo2012score, delong2011teamskill, makarov2017predicting}. 


\subsection{PreviousRank}

Elo, Glicko, and TrueSkill have several similarities in how they model and update a player's skill.
To provide a naive baseline we describe PreviousRank.

PreviousRank simply assumes that a player's predicted rank is equal to their observed rank in their previous match.
If a player is new to the system, we assume that their predicted rank is equal to $\frac{N}{2}$ where \textit{N} is the number of players competing in the match.
We use PreviousRank as our naive baseline to achieve a better understanding of the predictive power of other mainstream models.


\subsection{Calculating Predicted Ranks}

Elo, Glicko, TrueSkill, and our naive baseline all maintain a number that can be interpreted as the skill level of a player.
This estimation of the player's skill can be updated after every match.
These matches might be head-to-head or larger free-for-all matches.
Given such ratings, we can predict the ranking of a player in a field of other players.

For an upcoming match, we collect the players in the match.
For each player, we retrieve their rating.
If the player is new, we use the default rating value.
Players are sorted by their ratings thereby producing a rank prediction for the list of players.
Ties in the ratings are randomly broken.
By comparing this pre-match predicted ranking to the post-match observed ranking we can evaluate the performance of the rating systems.


\section{Metrics}
\label{sec: metrics}

As shown in the previous section, a rating system can be used to produce a ranking for a field of players in a match.
Given the predicted rankings $R^{pred}$ and observed rankings $R^{obs}$ after the match is finished, several metrics can be used to evaluate the performance of a rating algorithm.

Traditional metrics such as accuracy, mean absolute error, and rank correlation coefficients are commonly used for evaluating head-to-head games, but may not be as appropriate in free-for-all games.
We describe these three metrics and leverage three additional metrics taken from the field of information retrieval: average precision, mean reciprocal rank, and normalized discounted cumulative gain.
These metrics may yield different insights into the rating systems they evaluate.
In the remainder of this section, we present these metrics.


\subsection{Accuracy}

The problem of ranking may be viewed as a classification problem.
In a head-to-head match, accuracy can be used as the evaluation metric for a ranking problem with only two outputs or labels (three if a draw is possible).
In this case, each rank could be assumed as a nominal or categorical value.

Free-for-all matches can have many more players.
We could use the same assumption and treat the problem as a multi-label classification problem, evaluating the rankings based on how accurately they classify players into their observed ranks.
As such, accuracy is calculated as the ratio of correctly classified ranks to the total number of players.

Accuracy is not generally suited to ranking problems because it treats all the ranks as labels.
If a player was predicted to achieve rank 5 and earns rank 5, that is a hit.
But if he earns rank 6 or rank 96, it is a miss even when these two scenarios differ greatly.


\subsection{Mean Absolute Error}

Mean absolute error (MAE) is one of the most common measures to evaluate the similarity of two sets of values.
Assuming two rankings, the predicted ranks $R^{pred}$ and the observed ranks $R^{obs}$ of players, MAE is the average of the absolute errors:

\small
\begin{equation*}
    MAE = \frac{1}{N} \sum_{i=1}^{N}|R^{pred}_i - R^{obs}_i|    
\end{equation*}
\normalsize

\noindent where N is the total number of players competing in a match and $R^{pred}_i$ and $R^{obs}_i$ are the predicted and observed rankings for a player $p_i$.
An MAE of zero means that the two rankings are identical.
A higher MAE suggests higher dissimilarities between the two rankings.

MAE is more suited than accuracy to compare two sets of non-ordinal values.
Unlike accuracy, the case of a player having the predicted rank of 5 while earning the 96\textsuperscript{th} rank will have a much larger impact on the metric than if the player earned the 6\textsuperscript{th} rank.

However, MAE misses potentially useful information.
It does not distinguish between the prediction errors in higher ranks and those in lower ranks.
For example, MAE treats the difference between rank 1 and rank 6 the same as the difference between rank 90 and rank 95.
Accuracy suffers from the same issue.


\subsection{Kendall's Rank Correlation Coefficient}

Kendall's rank correlation coefficient, referred to as Kendall's tau and denoted by $\tau$, is a common statistic to measure the ordinal association between two variables~\cite{kendall1948rank}.
Kendall's tau leverages a more interpretable approach compared to other rank correlation coefficients by looking at the number of concordant and discordant pairs of observations.

Assume two rankings of $R^{pred}$ and $R^{obs}$ as the predicted rank and observed rank of players in a match between \textit{N} players.
For two players, $p_i$ and $p_j$, any pair of observations $(R^{pred}_i, R^{obs}_i)$ and $(R^{pred}_j, R^{obs}_j)$ are concordant if $R^{pred}_i > R^{pred}_j$ and $R^{obs}_i > R^{obs}_j$, or $R^{pred}_i < R^{pred}_j$ and $R^{obs}_i < R^{obs}_j$.
Otherwise, they are considered discordant.

Kendall's tau can be calculated as:

\begin{equation*}
    \tau = \frac{n_c - n_d}   {\binom{N}{2}}
\end{equation*}

\noindent where $n_c$ and $n_d$ are the number of concordant and discordant pairs.
The denominator is the total number of pair combinations.
Tau is equal to 1 if the predicted and observed rankings completely agree, is equal to -1 if they completely disagree, and is zero if there is no correlation between the two rankings.

Kendall's tau, unlike accuracy and MAE, does not consider the deviation in predicted and observed rankings, but instead considers the pairwise agreement between two rankings.
For example, if two players were predicted to have ranks 5 and 10 and achieved ranks 3 and 12, tau considers this a concordant pair without regard to the deviations in predicted versus observed ranks.
However, like accuracy and MAE, it does not distinguish between higher rank and lower rank errors.


\subsection{Mean Reciprocal Rank}

As the first metric adapted from the field of information retrieval, we leveraged mean reciprocal rank (MRR)~\cite{voorhees1999trec}.
It is often used to evaluate the performance of a query-response system that returns a ranked list based on a query.

We extend MRR to the evaluation of rank prediction in online competitive games.
Given the predicted ranks, $R^{pred}$, and observed ranks, $R^{obs}$, we compute the error for each player as the absolute difference between his predicted rank and observed rank.
In a free-for-all match with \textit{N} players, MRR can be calculated as:

\small
\begin{equation*}
    MRR = \frac{1}{N}\sum_{i=1}^{N}\frac{1}{1 + error_i}
\end{equation*}
\normalsize

\noindent where $error_i$ is the error in prediction.
The fraction of $\frac{1}{1 + error_i}$ may be considered as the relevance of the prediction for player $p_i$.
When the prediction is perfect the fraction is 1; the worse the prediction, the closer to 0 it becomes.
Therefore, MRR can be considered as a summation of relevance scores.

While the modified MRR applies a different penalty function than MAE, it is similar in that it considers higher penalties for higher differences between predicted and observed ranks.
However, like the above metrics, it considers the deviation between rank 1 and rank 6 to be the same as the difference between rank 90 and rank 95.


\subsection{Average Precision}

Average precision (AP) is the second metric we borrow from the field of information retrieval.
Given a ranked list of generated responses for a query, AP uses list-wise precision and relevance scores to evaluate the system~\cite{salton1991developments}.

We extend AP to the evaluation of rank prediction in online competitive games.
Similar to MRR, we consider $\frac{1}{1 + error_i}$ as the relevance score of each prediction.
In a free-for-all match with \textit{N} players, AP can be calculated as:

\small
\begin{equation*}
    AP = \frac{1}{N}\sum_{i=1}^{N} P(i) \times \frac{1}{1 + error_i}
\end{equation*}
\normalsize

\noindent where \textit{P(i)} is the overall precision value up to the \textit{i}\textsuperscript{th} position and $error_i$ is the error in prediction.

AP works exactly like MRR if all predictions are correct.
However, AP is generally more strict since it weights relevance scores for each position with the overall precision value up to that position.
Using precision values as weights causes AP to distinguish between prediction errors in higher ranks and lower ranks.
However, it puts a higher concentration on hits, especially in higher ranks.
This may have negative impacts on the evaluation of a model whose overall performance is great but its first few incorrect predictions occur in higher ranks, even when the prediction error is as small as 1 rank.
AP may be an appropriate metric for evaluating systems whose main focus is on high-rank or top-tier players.


\begin{figure*}
   \centering
\begin{tabular}{>{\centering\arraybackslash} m{0.7cm} >{\centering\arraybackslash} m{3.77cm} >{\centering\arraybackslash} m{3.4cm} >{\centering\arraybackslash} m{3.4cm} >{\centering\arraybackslash} m{3.4cm} }
\hline
Metric & All Players & Best Players & Frequent Players & Binned Ranks\\
\hline
\rotatebox[origin=c]{90}{Accuracy}&
\includegraphics[width=4.1cm]{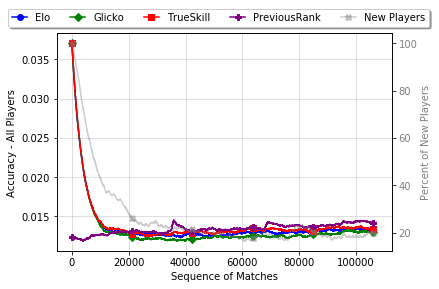}&
\includegraphics[width=3.7cm]{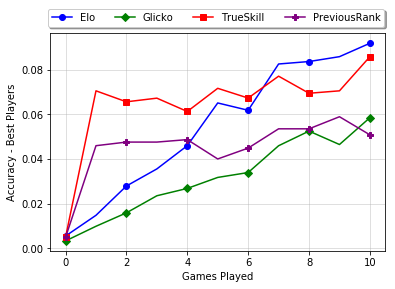}&
\includegraphics[width=3.7cm]{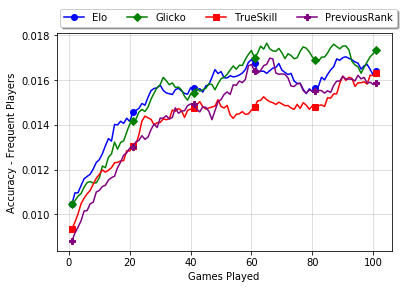}&
\includegraphics[width=3.77cm]{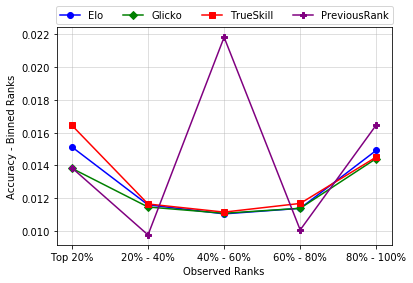}\\
\rotatebox[origin=c]{90}{MAE}&
\includegraphics[width=4.1cm]{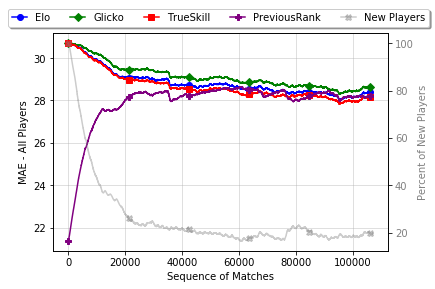}&
\includegraphics[width=3.7cm]{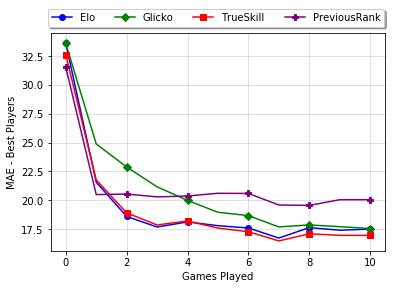}&
\includegraphics[width=3.7cm]{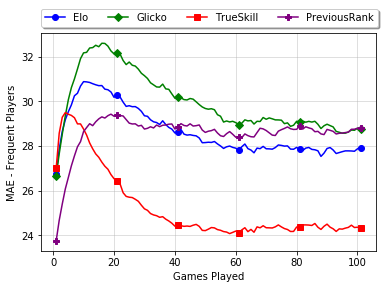}&
\includegraphics[width=3.77cm]{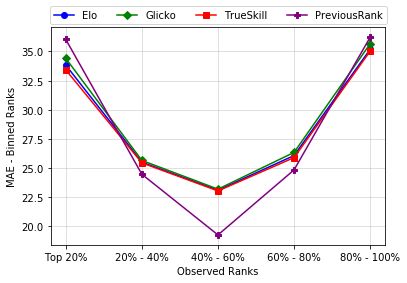}\\
\rotatebox[origin=c]{90}{Tau}&
\includegraphics[width=4.1cm]{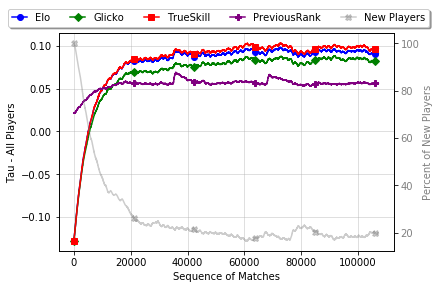}&
\includegraphics[width=3.7cm]{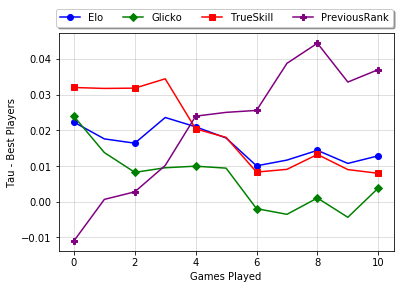}&
\includegraphics[width=3.7cm]{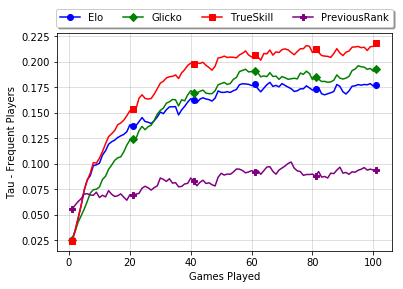}&
\includegraphics[width=3.77cm]{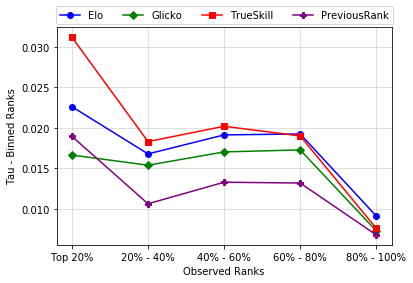}\\
\rotatebox[origin=c]{90}{MRR}&
\includegraphics[width=4.1cm]{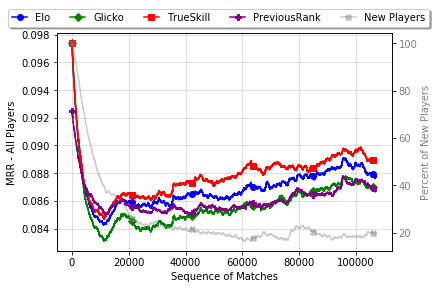}&
\includegraphics[width=3.7cm]{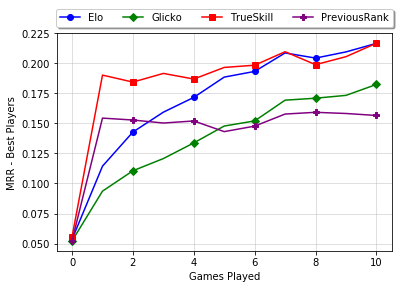}&
\includegraphics[width=3.7cm]{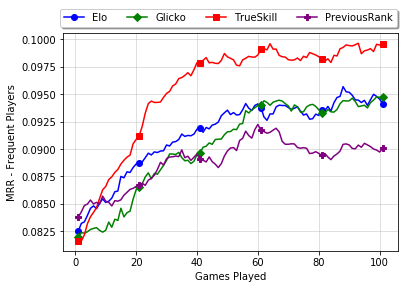}&
\includegraphics[width=3.77cm]{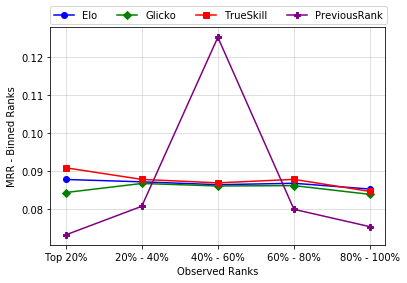}\\
\rotatebox[origin=c]{90}{AP}&
\includegraphics[width=4.1cm]{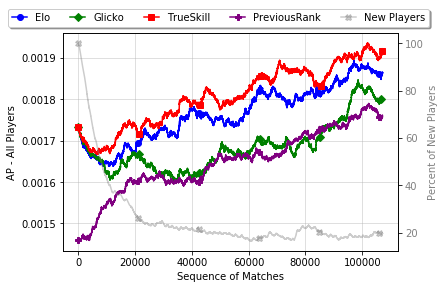}&
\includegraphics[width=3.7cm]{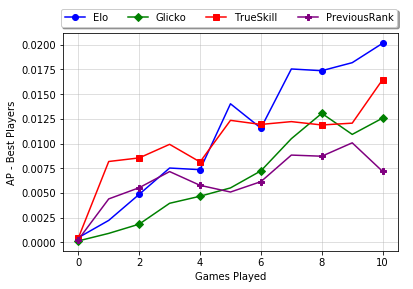}&
\includegraphics[width=3.7cm]{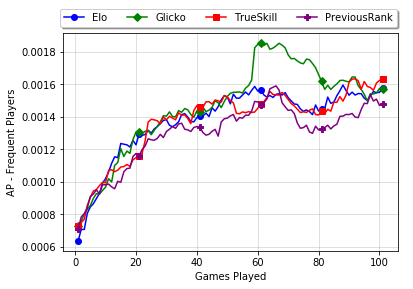}&
\includegraphics[width=3.77cm]{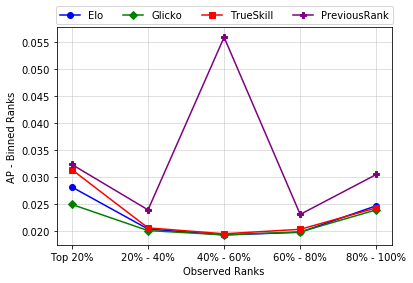}\\
\rotatebox[origin=c]{90}{NDCG}&
\includegraphics[width=4.1cm]{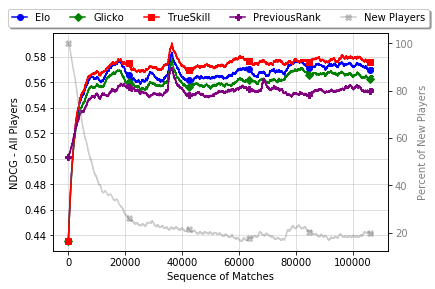}&
\includegraphics[width=3.7cm]{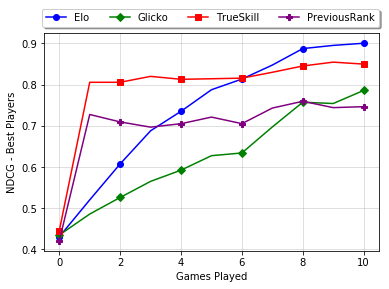}&
\includegraphics[width=3.7cm]{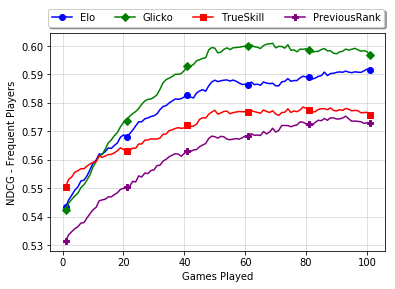}&
\includegraphics[width=3.77cm]{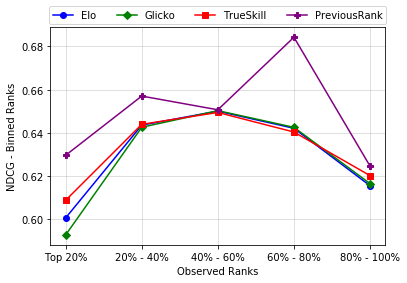}\\
\end{tabular}

\caption{The results of evaluating Elo, Glicko, TrueSkill, and PreviousRank using accuracy, MAE, Kendall's tau, MRR, AP, and NDCG on four different experimental set-ups: all players, best players, frequent players, and binned ranks.}
\label{fig:results} 
\end{figure*}


\subsection{Normalized Discounted Cumulative Gain}
Normalized Discounted Cumulative Gain (NDCG) is the third metric we adapt from the field of information retrieval.
Given a ranked list of responses for a query, NDCG evaluates the quality of the generated responses based on their relevance score and weighted position in the list~\cite{jarvelin2002cumulated}.
The overall score is accumulated from individual scores at each level from the top of the list to the bottom.

We extend NDCG to the evaluation of rank prediction in online competitive games.
Similar to MRR and AP, we consider $\frac{1}{1 + error_i}$ as the relevance score of predictions.
In a free-for-all match with \textit{N} players, NDCG can be computed as:

\begin{equation*}
    NDCG = \frac{\sum_{i=1}^{N} \frac{1}{log_{2}(i+1)} \times \frac{1}{1+error_i}    }{IDCG}
\end{equation*}

\noindent where $error_i$ is the prediction error, $\frac{1}{log(i+1)}$ is the weight assigned to the \textit{i}\textsuperscript{th} position, and \textit{IDCG} is a normalizing factor.

Similar to AP, NDCG distinguishes between errors in higher ranks and lower ranks by weighting the evaluation of each position.
However, in NDCG we can adjust the weights based on the evaluation goals.
For example, by increasing the weights, the model may direct its attention to evaluating good or regular players who often appear in higher ranks.


\section{Methodology}
\label{sec: methodology}

In this section, we introduce the dataset used to perform our experiments.
We detail our methodology along with the parameters we used to implement the rating systems.
Finally, we explain how we performed our evaluations.

PlayerUnknown's Battlegrounds (PUBG) is a popular free-for-all online multiplayer video game developed and published by PUBG Corporation.
The game pits up to one hundred players in a battle royale match against each other.
The last player or team standing wins the match.
PUBG is played in teams of four, teams of two, or singletons.
The dataset is publicly available on \textit{\url{www.kaggle.com}}, a public data platform.

For this research, we only considered solo matches in the dataset; free-for-all matches where each player competes against every other player at the same time.
In this paper, we are concerned with how to evaluate a single player, a necessary step before we can evaluate a team.
The dataset provides in-game statistics such as the number of kills, distance walked, and rank for over 100,000 unique matches and 2,260,000 unique players.

We sorted the matches by their timestamps.
For each match, in order, we retrieved the list of players.
New players -- those that have yet to appear in a match -- were assigned default ratings, 1500 for Elo and Glicko, and 25 for TrueSkill.
We also retrieved the skill rating of returning players.

Based on these skill ratings, we sorted the players and used this sorting as the rank prediction for the match.
The ratings for the players were updated after each match by comparing their predicted ranks with their observed ranks.
The parameters we used for each rating system include \textit{k = 10} and \textit{D = 400} for Elo, and $\beta$ = 4.16 and $\tau$ = 0.833 for TrueSkill as suggested in its official documentation.

For each match, the metrics (accuracy, MAE, etc.) were computed creating a time-series reporting the performance of the rating systems.
To aid in our exploration of the evaluation metrics we implemented four different set-ups.

First, we evaluated the performance of the rating systems for all players and all unique matches in the dataset sorted by date.
In this set-up, every player is treated equally regardless of their skill or how often they play.

Second, we evaluated the performance of the rating systems for the best players in the system.
To identify the best players, we selected the 1000 players with the highest ratings who had played more than 10 games.
Since these players did not compete at the same time, we evaluated the predictive performance of the rating systems on their first 10 games.

Third, we evaluated the performance of the rating systems for the most frequent players in the system.
To identify the most frequent players, we randomly selected 1000 players who played more than 100 games.
We evaluated the predictive performance of the rating systems on their first 100 games.

Finally, we evaluated the performance of the rating systems for binned ranks -- a grouping of players based on their observed ranks.
for each individual match, we divided the competing players into five different bins.
The top 20\% players of each match may be considered as skilled players while the last 20\% may be viewed as novice players.
The three middle bins may contain seasonal players or players who are still learning the game and advancing their skills.
New players may negatively influence the system's performance in this set-up.
Almost half of the players in the dataset only played one game.
The rating systems do not have any knowledge about these players and yet they can be placed in any of the five bins.
For each evaluation metric, we averaged the score of each rating system for each bin over all matches.


\section{Results and Discussions}
\label{sec: results}

In this section, we discuss the results of four experiments on all players, best players, most frequent players, and binned ranks.
We compare the performance of the competing models using six evaluation metrics discussed earlier.
Finally, we analyze the ability of these metrics in capturing prediction patterns of the rating systems.

The results of these experiments are given in Figure~\ref{fig:results}.
In this figure, the rows correspond to evaluation metrics and the columns are associated with experimental set-ups.
The performance of the models is represented by trend lines with different colors.
For example, in accuracy plot for the best players, the accuracy of TrueSkill, shown by the red trend line, starts from 0.5\% and increases up to 8.5\% after 10 games.


\subsection{Accuracy}

Previous works demonstrated that Elo, Glicko, and TrueSkill achieve high accuracy for predicting ranks in head-to-head games.
However, the results shown in figure~\ref{fig:results} indicate that it does not hold true for free-for-all games.

The observed accuracy values are fairly small in all set-ups.
The highest value is around 10\% in the case of Elo for the best players.
In addition, the results display different patterns.
For example, accuracy suggests that PreviousRank, Elo, and Glicko have relatively better performance for evaluating all players, best players, and frequent players, respectively.

Accuracy seems to be influenced by factors such as the number of new players, players' behavior, and their frequency of play.
We expect the rating systems to achieve a better knowledge of players' skills over time by observing more games.
For all players, Elo, Glicko, and TrueSkill achieved higher accuracy than PreviousRank at the early stages of the sequence of matches where most of the players are newly added to the system (showed by the gray trend line).
These matches contain all or many ties in ratings that are randomly broken by the rating systems.
The patterns show that the rating systems cannot outperform random predictions of the early stages of the sequence.
Accuracy values significantly drop as the number of known players to the system increases.
However, the patterns observed for the best and frequent players suggest the opposite when the influence of new players is excluded.
The ratings converge faster for the best players -- the players whom we expect to show consistent behaviors.

Results of binned ranks reveal that the models can correctly distinguish between different types of players based on accuracy.
Elo, Glicko, and TrueSkill achieved higher accuracy for players in the first and last bins who are assumed to demonstrate consistent behavior (highly skilled players and novice players).
On the other hand, the models achieved less accuracy for the middle bins that presumably contain players with inconsistent behavior (less known or seasonal players).

Accuracy is a reliable metric for evaluating head-to-head games.
However, as the results suggest, it is not suited to evaluating free-for-all games. 
Accuracy highly exaggerates the negative influence of new players on the performance of rating systems.
It also treats all the ranks as labels and thus, is unable to explain the real differences between two ranks.
Finally, since accuracy only considers hits as good predictions, it is unable to capture a large part of the predictive behavior of rating systems.


\subsection{MAE}

The results of evaluating rating systems using MAE suggest significant errors for predicting rank in free-for-all games.
For example, the lowest MAE value observed was 17 for TrueSkill in the case of the best players.

The results display fairly similar patterns for Elo, Glicko, and TrueSkill in all, best, and frequent players set-ups.
For example, TrueSkill achieved the lowest MAE values while Glicko experienced higher errors in all these three set-ups.

While accuracy was mostly influenced by the number of new players, MAE seems fairly resilient in comparison.
Elo, Glicko, and TrueSkill achieved a better knowledge of players improving with a slow rate as the number of new players in each match decreases over time.
On the other hand, MAE seems to be highly influenced by players' behavior.
The models demonstrate considerable improvements by observing more games from the best players.
However, for frequent players set-up that includes players with different skills, the patterns are not as clear.

Finally, the MAE plot for binned ranks suggests that the rating systems achieved the lowest errors for the three middle bins (i.e. players with inconsistent behaviors or less-known players) while achieving higher errors for the first and last bins (i.e. players with consistent behaviors or known players).

While MAE is more suited than accuracy for evaluating rank predictions, the results suggest that it is not suited for evaluating free-for-all games.
MAE presents a global evaluation of the model's performance.
However, since MAE does not consider rank positions, it cannot explain the real differences between players and fails to capture the evaluation details.


\subsection{Kendall's Tau}

Similar to accuracy, Kendall's tau seems to amplify the influence of new players.
Elo, Glicko, and TrueSkill show negative correlations at the early stages of the sequence in all players set-up.
However, the correlations increase with a fast rate as the number of new players in each match decreases.

The patterns observed for the best players suggest that Kendall's tau is unable to capture the learning ability of the rating systems from players with consistent behavior.
Elo, Glicko, and TrueSkill demonstrate decreasing correlation values as they observe more games while PreviousRank shows an upward trend.
On the other hand, the frequency of play seems to be an important factor influencing the evaluations of Kendall's tau.
The results of the frequent players set-up suggest that, regardless of players' skill levels, Kendall's tau captures the ability of rating systems in achieving a better knowledge of players by observing more games.
The learning demonstrated by Kendall's tau for frequent players happens much faster than what we observed for accuracy and MAE.
For example, TrueSkill starts from 2.5\% correlation while constantly increasing to reach 22\% correlation at the end of the 100\textsuperscript{th} game.

Finally, the patterns observed in the binned ranks plot imply that based on Kendall's tau, Elo, TrueSkill, and Previous Rank perform best for the first bin which consists of highly skilled players.
This is inconsistent with the patterns Kendall's tau showed for the best players set-up.

The results suggest that Kendall's tau is not suited to evaluating rank predictions in free-for-all games.
It cannot represent the real predictive power of rating systems under the influence of new players.
It also fails to capture the learning patterns of the systems for top-tier players over time.
Kendall’s tau only considers the pairwise agreement between predicted and observed ranks without regard to their deviations.
Therefore, it cannot capture the real differences between players.


\subsection{MRR}

MRR correctly captured the ability of rating systems in achieving more knowledge of players by observing more games based on their playing behavior and frequency of play.
The models show different learning patterns.
The fastest learner is TrueSkill in both best and frequent players set-ups.
MRR also suggests that TrueSkill is the best performer overall. 

Results of MRR show fairly similar patterns to accuracy.
However, MRR values are higher than accuracy for all set-ups.
In addition, MRR seems to be more resilient than accuracy to the influence of new players.
Accuracy required at least 80\% of players in the match to be known to the system in order to start its upward trend after the early drop while this value for MRR is around 60\%.
All models also learn faster based on MRR compared to what we observed for accuracy.

Finally, based on the results of MRR for binned ranks, Elo and TrueSkill show a fairly similar pattern to that of Kendall's tau.
Elo and TrueSkill performed their best predictions for the first bin while showing higher errors for the last bin.
On the other hand, Glicko and PreviousRank show the exact opposite of what we expected, higher MRR values for the middle bins, and lower MRR values for the first and last bins.

Although MRR is a generic rank-based evaluation metric (like Kendall's tau), it relatively captured the ability of rating systems to learn the player's behavior over time.
However, it is highly influenced by the number of new players in matches.
It also does not distinguish between higher rank and lower rank prediction errors.
Using metrics that adjust their evaluation scores based on the prediction error's position may better benefit the evaluation by giving us the flexibility to adjust the evaluations based on our goals.


\subsection{Average Precision}

AP patterns suggest that the models learn more about the players over time by observing more games.
This is particularly true for players with consistent behavior and players who play frequently.
However, AP suggests a fairly slow learning process for all models.

While accuracy, Kendall's tau, and MRR were substantially influenced by the number of new players and changed dramatically over the first few matches in all players set-up, the models experience a small drop in AP values for the same matches and start improving with a fairly fast rate afterward. 
This pattern demonstrates that AP is more resistant to the influence of new players in the system.
For example, TrueSkill corrected its early downward trend when at least 45\% of the players in a match are known to the system.

Finally, AP results for binned ranks show that Elo, Glicko, and TrueSkill achieved higher AP for players who are assumed to be known to the system (the first and last bins) while displaying more uncertainties for less-known players (the middle bins).
This pattern is fairly similar to that for accuracy.

AP demonstrated three main benefits over previous metrics.
First, it is more resistant to the influence of new players.
Second, it considers rank positions when evaluating predictions and thus, can capture the differences between players.
Finally, it more accurately makes predictions for the four experimental set-ups.
However, AP values are extremely small.
AP provides strict evaluations of the models by using precision values as weights for the relevance of each prediction, corresponding to the relative importance of each position.
These weights may over-penalize trivial errors that are not considered bad predictions in other metrics.


\subsection{NDCG}

NDCG accentuates the high learning ability of the rating systems based on the consistent behavior of players and their frequency of play.
For example, in the best players set-up, TrueSkill shows a significant increase from 45\% to 80\% after observing one game. 
The patterns show that Elo is the best performer for the best players achieving NDCG of 90\% at the end of the 10\textsuperscript{th} game.
On the other hand, Glicko shows the best performance for the most frequent players.

NDCG seems to be the metric least affected by the influence of new players in the system.
It starts increasing from the start and improves with a highly fast rate as the number of new players in matches decreases.
The learning process gradually becomes slower as the number of new players in each match does not change much.

Finally, although NDCG showed good performance in capturing the predictive performance of the rating systems, the patterns observed for binned ranks contradict our expectations.
The models achieved their best performance for the middle bins while achieving lower scores for the first and last bins.
As mentioned before, this could well be the result of the large number of new players who may be placed in either of bins. 

NDCG alleviates all the challenges faced by other metrics.
The weighting factor used in NDCG is separate from the predictions and is directly based on the positions.
Therefore, before evaluating the predictions, we can adjust the weights based on the goals of the system.
The results suggest that NDCG correctly captures the learning ability of the rating systems for both players' behavior and frequency of play.
It can also represent the predictive power of the rating systems even when a large number of players in a match are new to the system.
Finally, NDCG seems to be the best metric for evaluating rank predictions in free-for-all games.


\section{Conclusion and Future Works}
\label{sec: conclusion}

In this paper, we evaluated the predictive performance of three popular rating systems in free-for-all games.
We performed our experiments on four different groups of data to paint a clear picture of the evaluations.

The results indicated that many metrics were negatively influenced by the number of new players in each match. 
Some metrics captured the ability of rating systems to learn more about the behavior of players by observing more games.
Others correctly captured the differences between players.
some metrics while being well suited to evaluate the rating systems on a certain group of players, may not be appropriate for other groups of players.
Achieving better predictions for Top-tier players is particularly more important since these players often stay in the system and play more games.

Among all metrics tested, NDCG best represented the predictive power of rating systems while resolving all challenges faced by other metrics.
It was more resistant than other metrics to the influence of new players.
It also correctly captured the learning patterns of these systems based on both player's behavior and frequency of play.

Our experimentation was limited to free-for-all games.
Evaluating other modes of game-play is part of our future work.
This work is a part of more comprehensive research on group assignment in online competitive games.
Evaluating rank predictions is the first step in building a framework for predicting rank.
We plan to extend rating systems by incorporating players' behavioral features to achieve better predictions.
We will extend rank prediction to building a more comprehensive framework for predicting the success of proposed teams and making assignments.


\bibliographystyle{IEEEtran}
\bibliography{IEEEabrv, main}

\end{document}